# Possible Existence of Partially Disordered Sm Ions in Magnetically Ordered State of Ising Magnet $SmPt_2Si_2$: a Single Crystal Study

Kengo Fushiya[1]*, Tatsuma D. Matsuda[1], Ryuji Higashinaka[1], Kazuhiro Akiyama[2], and Yuji Aoki[1]

[1]*Department of Physics, Tokyo Metropolitan University, Hachioji-city, Tokyo 192-0397, Japan*

[2]*Department of Chemistry, Tokyo Metropolitan University, Hachioji-city, Tokyo 192-0397, Japan*

We have succeeded in growing single crystals of $SmPt_2Si_2$ and have performed electrical resistivity, magnetization and specific heat measurements. The magnetic susceptibility shows that $SmPt_2Si_2$ is an Ising magnet with the crystalline-electric-field ground state of $J_z=\pm 3/2$. We have found the existence of two magnetically ordered states, i.e., an antiferromagnetically (AFM) ordered phase (I) setting in at $T_N$=5.1 K and a field-induced magnetization plateau phase (II). In the phase I, unlike the usual AFM states, a pronounced Curie-Weiss contribution remains in magnetic susceptibility, indicating the existence of "partially disordered" Sm ions. Largely enhanced Sommerfeld-coefficient (350 mJ/K$^2$mol) in the phase I can be attributed to formation of Kondo sublattice with heavy quasiparticles in the partially disordered Sm ions. These findings may reflect substantial magnetic frustrations inherent in $SmPt_2Si_2$.

In this decade, there has been increasing interest in the strongly correlated electron behaviors appearing in Sm-based intermetallic compounds. Typical such examples are unusual magnetic-field-insensitive heavy-fermion behavior in $SmOs_4Sb_{12}$,[1,2] magnetic-field-insensitive phase transitions and largely-enhanced Sommerfeld coefficients in $SmT_2Al_{20}$ ($T$=Ti, V, Cr, and Ta),[3,4,5] metal-insulator (MI) transition and magnetic-field-induced charge ordering in $SmRu_4P_{12}$,[6,7,8] and MI transition under pressure in Sm$X$ ($X$ = S, Se, and Te).[9,10,11] Since Sm ions in all of these compounds have cubic site symmetries, the crystalline-electric-field (CEF) ground states tend to have high degeneracy with active multipole degrees of freedom. In order to investigate the role of such CEF degeneracy on the strongly correlated electron behaviors, it may be useful to study "anisotropic" systems and compare them with the cubic ones.

One of the appropriate candidate materials with such anisotropy is $RT_2X_2$ ($R$: rare earth and actinide, $T$: transition metal, $X$: Si and Ge) with a tetragonal crystal structure. In the Ce-, Yb- and U-based compounds, *e.g.*, $CeCu_2Si_2$,[12,13] $YbRh_2Si_2$,[14,15] and $URu_2Si_2$,[16,17,18] a variety of characteristic features including heavy-fermion behavior, non-Fermi liquid behavior and anisotropic superconductivity appear. For Sm based compounds, however, only a few reports have been made of systematic investigations of the physical properties using single crystals.[19,20]

In this paper, we report the first successful single crystal growth of $SmPt_2Si_2$ and measurements of electrical resistivity, magnetization and specific heat. We have found that $SmPt_2Si_2$ has a strong Ising anisotropy and exhibits multiple magnetically ordered states at low temperatures. Largely enhanced specific heat in the ordered state indicates existence of strong correlations in $SmPt_2Si_2$.

Single-crystals of $SmPt_2Si_2$ were grown by the Sn-flux method. The starting materials were 3N(99.9% pure)-Sm, 4N-Pt, 4N-Si, and 5N-Sn. These materials were inserted in an alumina crucible with an off stoichiometric composition of Sm : Pt : Si : Sn=1 : 3 :1 : 40, and sealed in a quartz tube. The quartz tube was heated up to 1150 °C, maintained at this temperature for 1 day, and cooled down to 650 °C at a rate of -2 °C/h, taking about 8 days in total. Single crystals were obtained by spinning the ampoule in the centrifuge to remove excess Sn-flux. Typical single crystals are shown in Fig. 1.

The crystal structure of $SmPt_2Si_2$ has been refined by means of single-crystal X-ray diffraction analysis using a Rigaku Mercury diffractometer with graphite monochromated Mo-K$\alpha$ radiation. A selected small single crystal with dimensions of 0.1×0.07×0.05 mm$^3$ was mounted on a glass fiber with epoxy. We have confirmed that $SmPt_2Si_2$ has the primitive tetragonal $CaBe_2Ge_2$ type structure (Space Group: *P4/nmm*, #129) as shown in Fig.2.[21] Using the program SHELX-97,[22] we have determined for the first time the structural parameters, which are summarized in Table I. In the parameters of the respective site occupancies and *B*-factors and the *R*-factor, there is no noticeable evidence for an incorporation of Sn ions into the Sm, Pt and Si sites. An energy dispersive X-ray spectroscopy (EDX) analysis indicates that the upper bound of Sn incorporation is about 0.03 Sn-ion/f.u., even if it exists.

As shown in Fig. 2(b), the $CaBe_2Ge_2$ type structure is similar to the well-known $ThCr_2Si_2$ type structure.[23] The major difference is the Sm site symmetry (4*mm* for the $CaBe_2Ge_2$ type and 4/*mmm* for the $ThCr_2Si_2$ type), which is caused by the different stacking of *T* and *X* layers along the *c*-axis.

Transport properties were measured using a standard AC four-probe technique in a commercial Quantum Design (QD) physical property measurement system (PPMS). DC magnetization *M* was measured in a QD magnetic property measurement system (MPMS) down to 1.8 K and up to 7 T. Specific heat was measured by a quasi-adiabatic heat pulse method using the QD PPMS and a $^3$He–$^4$He dilution refrigerator down to 0.2 K.

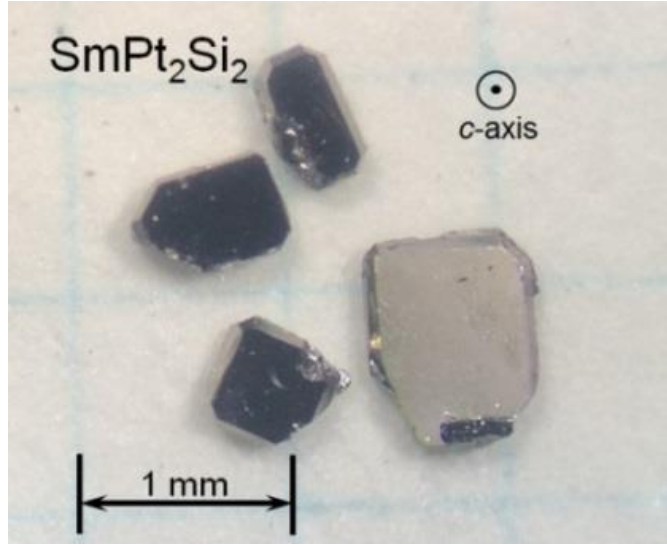

Fig.1 (Color Online) Typical single crystals of SmPt$_2$Si$_2$. The thickness is about 0.07~ 0.4 mm.

Table I. Atomic coordinates and thermal parameters of SmPt$_2$Si$_2$ at 300 K determined by single-crystal X-ray measurements (2$\theta_{max}$: 54.9 deg, reflections collected: 1265, independent reflections: 150). $R$ and $wR$ are the reliability factors and $B$ is the equivalent isotropic atomic displacement parameter.

| P4/nmm(#129) origin choice 1 | Atom (site) | x | y | z | B(Å$^2$) |
|---|---|---|---|---|---|
| a = 4.198 Å | Pt1 (2c) | 1/2 | 0 | 0.12217(8) | 0.87(6) |
| c = 9.818 Å | Pt2 (2a) | 0 | 0 | 1/2 | 0.29(6) |
| V = 173.1 Å$^3$ | Sm (2c) | 1/2 | 0 | 0.7545(8) | 0.39(6) |
|  | Si1 (2c) | 1/2 | 0 | 0.3648(87) | 0.3(2) |
|  | Si2 (2b) | 0 | 0 | 0 | 0.4(2) |
| (R= 4.7%, wR= 9.42%) | | | | | |

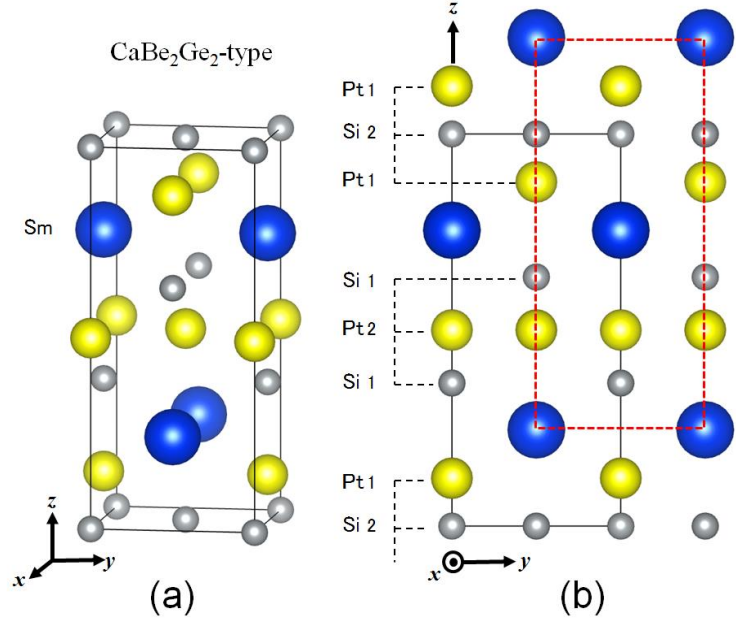

Fig.2 (Color Online) (a) Crystal structure of SmPt$_2$Si$_2$ and (b) the projection on the *bc* plane. The Sm-ion site does not have inversion symmetry.

Figure 3(a) shows the temperature dependence of magnetic susceptibility $\chi(=M/H)$ measured at 1 T. The most salient feature is pronounced anisotropy between *H//c* and *H//a*, clearly demonstrating an Ising anisotropy with the easy axis of magnetization along the tetragonal *c* axis. For *H//c*, a Curie-Weiss fitting using $\chi=C/(T-\theta_p)+\chi_0$ between 8 and 200 K provides $C$=0.080 emu K/mol, $\theta_p$=-0.46 K and $\chi_0$=-1.3x10$^{-4}$ emu/mol. The effective magnetic moment $\mu_{eff}$=0.80 $\mu_B$/Sm calculated from $C$ indicates the valence of Sm ions being close to 3+.

With the tetragonal site symmetry (4*mm*), the *J*=5/2 multiplet of Sm$^{3+}$ ions splits into three doublets due to the CEF effect represented by the Hamiltonian $\mathcal{H} = B_2^0 O_2^0 + B_4^0 O_4^0 + B_4^4 O_4^4$, where $B_i^j$ and $O_i^j$ are the CEF parameters and Steven's operators, respectively.[24] Among the doublets, only $|\pm\frac{5}{2}\rangle$ and $|\pm\frac{3}{2}\rangle$ (i.e., $d|\pm\frac{5}{2}\rangle+\sqrt{1-d^2}|\mp\frac{3}{2}\rangle$ with *d*=0 and 1) can reproduce the observed Ising behavior (no temperature dependence of $\chi$ for *H//a*). Comparing with the expected values of $C$ being 0.191 and 0.069 emu K/mol for $|\pm\frac{5}{2}\rangle$ and $|\pm\frac{3}{2}\rangle$, respectively, the observed $C$=0.080 emu K/mol indicates that the CEF ground state is most probably $|\pm\frac{3}{2}\rangle$.

In the $\chi(T)$ data, a small but clear cusp appears for *H//c* without noticeable anomaly for $H//a$, evidencing the existence of an antiferromagnetic (AF) transition occurring at $T_{N1}$=5.1 K. In a usual Ising magnet, $\chi$ should decrease below $T_{N1}$ approaching zero at $T$=0 for $H$ applied along the easy axis. In SmPt$_2$Si$_2$, however, $\chi$ turns into an increasing behavior below 3 K. This anomalous behavior indicates that the ordered state (phase I) in $T<T_{N1}$ is not a simple AF ordering.

Figure 3(b) shows the $T$ dependences of $M$ measured in various magnetic fields below 8 K for $H//c$. With increasing $H$, the cusp structure shifts to lower temperatures. Above 2 T, a high $M$ state appears at low temperatures and the boundary, where a sudden drop in $M$ appears at $T_{N2}$, shifts to higher temperatures with increasing magnetic field. Using the data shown in Fig. 3, we have constructed the $H$-vs-$T$ phase diagram for $H//c$ as shown in Fig.4.

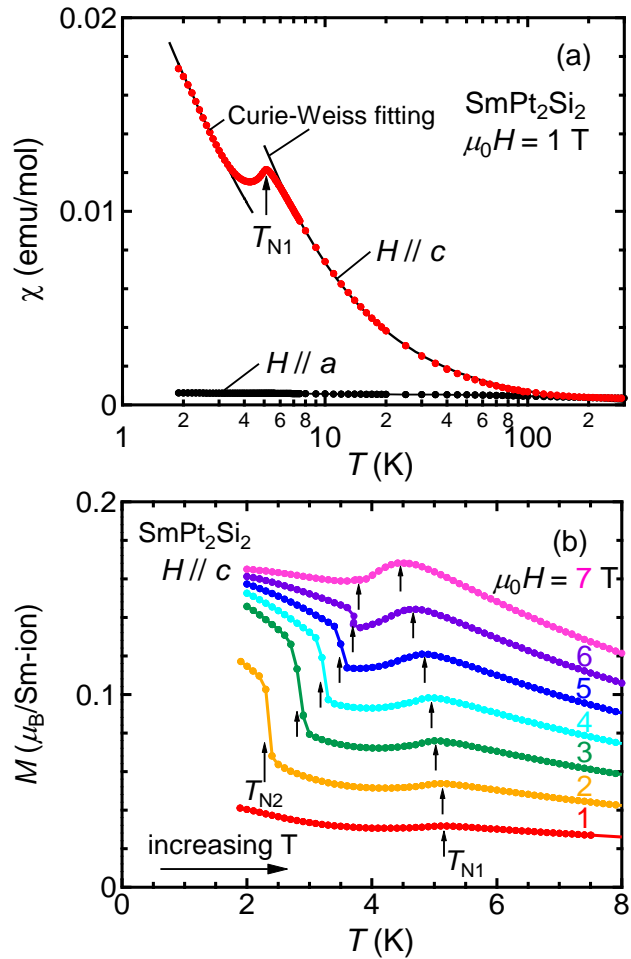

Fig.3 (Color Online) (a) Temperature dependence of magnetic susceptibility $\chi$ of SmPt$_2$Si$_2$ for $H//a$ and $H//c$. (b) Temperature dependences of magnetization measured in various magnetic fields for $H//c$.

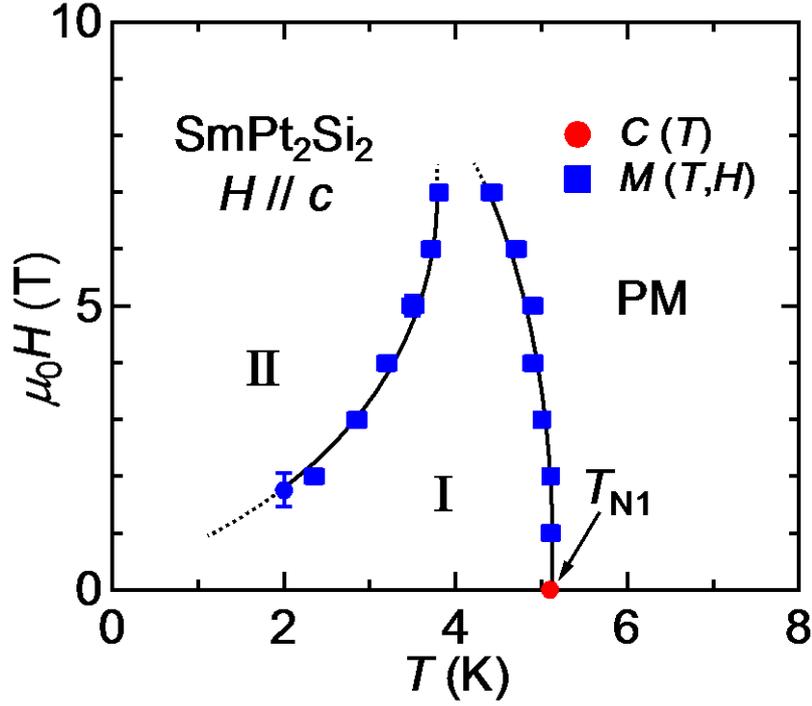

Fig.4 (Color Online) $H$ vs $T$ phase diagram of SmPt$_2$Si$_2$ for $H//c$. Hysteresis in the transition between phases I and II is shown by error bars.

Magnetization curves for $H//c$ and $H//a$ are shown in Fig.5. For $H//c$, a metamagnetic anomaly appears at the transition between phases I and II. The associated hysteretic behavior reflects the first-order nature of the transition. As the temperature increases, the field $H_M$ of the jump shifts to higher fields. In $H>H_M$ (phase II), $M$ shows weak $H$ dependence with $M\sim0.15$ $\mu_B$/Sm. This value is close to 1/3 of 0.45 $\mu_B$/Sm, which is the expected saturation value of $M$ for $|\pm\frac{3}{2}>$. Therefore, we simply expect that the magnetic structure in the phase II is ↑↑↓.

In phase I, $M$ in $H//c$ is almost proportional to the applied field and it has almost the same magnitude as in the paramagnetic state (see Fig. 5). This behavior is quite unusual; normally, $M$ of an Ising magnet for $H$ applied along the easy direction decreases sharply below the transition temperature and approaches zero. This anomaly strongly indicates that a large amount of disordered Sm ions still remains in the phase I.

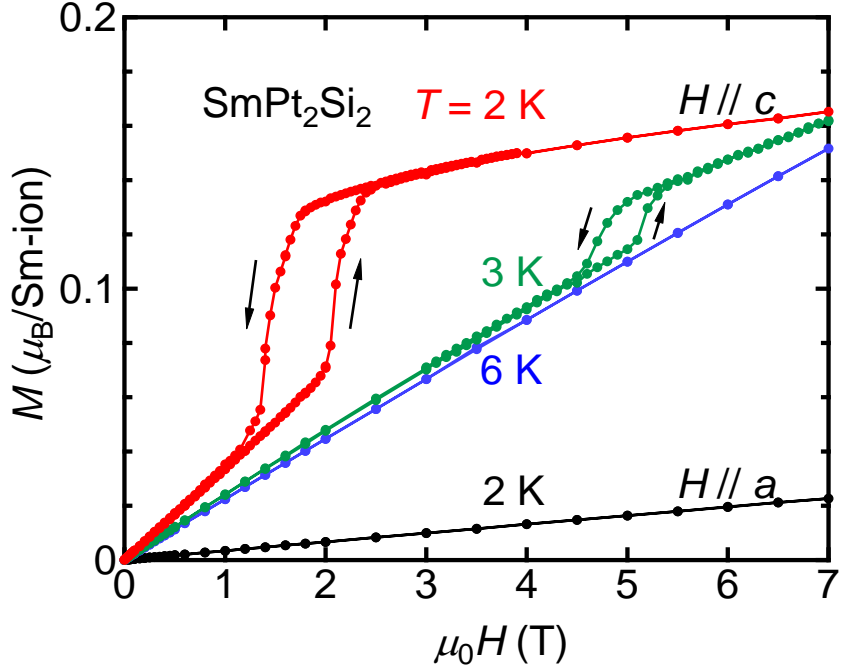

Fig.5 (Color Online) $H$ dependence of magnetization at low temperatures below 6 K.

Figure 6(a) shows the temperature dependence of specific heat $C/T$ down to 0.2 K in zero field. A pronounced λ-type anomaly appearing at $T_{N1}$ corresponds to the AFM transition. At $T^* \approx 1.8$ K, there appears a broad peak. Considering the broadness of the peak, we attribute this anomaly not to the existence of a phase transition but to some thermal excitations in the 4f-electron magnetic system in the phase I.

An upturn appearing below 0.5 K is due to the nuclear Schottky contribution, which can usually be expressed as $C_n = A_n/T^2$. Below 1 K, the $C$ data can be fitted using

$$C/T = \gamma + A_n/T^3 + \beta T^2 + \alpha T^n e^{-\Delta/T}, \qquad (1)$$

where the phonon contribution $\beta = 0.5$ mJ/(K$^4$·mol) is determined for LaPt$_2$Si$_2$[25] and the last term is phenomenologically introduced to express the $T$-dependent part of the 4f magnetic excitations. A fitting to the data below 1 K using Eq. (1) gives $\gamma \cong 350$ mJ/mol·K$^2$ and $A_n \cong 5.6$ mJ·K/mol. Comparing with $\gamma_{La} = 4$ mJ/(K$^2$·mol) for LaPt$_2$Si$_2$,[25] the large enhancement of the $\gamma$ value in SmPt$_2$Si$_2$ suggests the formation of heavy quasiparticles even in the ordered state of phase I.

Figure 6(b) shows the magnetic entropy $S_{4f}$ calculated using the data of $C_{4f} \equiv C - C_n - \beta T^3$.

The sizable release of the magnetic entropy is consistent with the valence of Sm ions being close to +3. Furthermore, above $T_{N1}$, $S_{4f}$ tends to saturate and approaches $R\ln 2$. This behavior agrees well with the above-mentioned CEF model, in which the ground state is a $\left|\pm\frac{3}{2}\right>$ doublet. The value of $S_{4f}$ at $T_{N1}$ is 80% of $R\ln 2$. This suppression in $S_{4f}$ can be attributed to the Kondo effect and/or short-range magnetic correlations developing in Sm 4f magnetic moments above $T_{N1}$.

The temperature dependence of electrical resistivity ρ for the current $J//a$ is shown in Fig. 7. At high temperatures, the resistivity shows a metallic behavior without any anomaly indicating a charge-density-wave transition as observed in $LaPt_2Si_2$ [26]. Below $T_{N1}$, ρ starts to increase abruptly and tends to saturate approaching $T=0$. This behavior is probably due to the superzone gap opening associated with the AFM ordering; similar resistivity anomalies have been also observed in $URu_2Si_2$ [17] and Cr [27]. The relatively small value of $(\rho(T\to 0)-\rho(T_{N1}))/\rho(T_{N1})=0.025$ indicates that only small part of the Fermi surface disappears on entering the phase I.

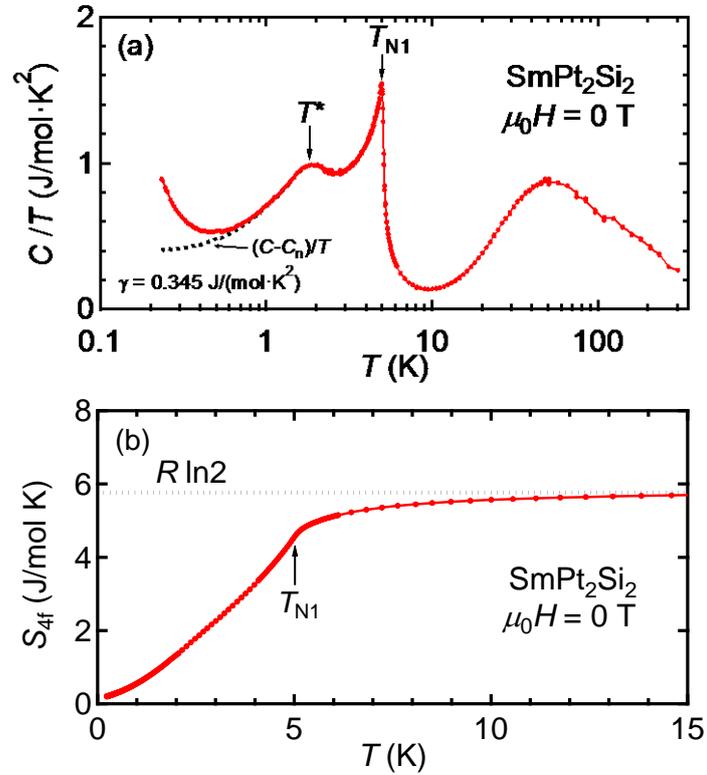

Fig.6 (Color Online) $T$ dependence of specific heat $C/T$ (a) and magnetic entropy $S_{4f}$ (b) in zero field.

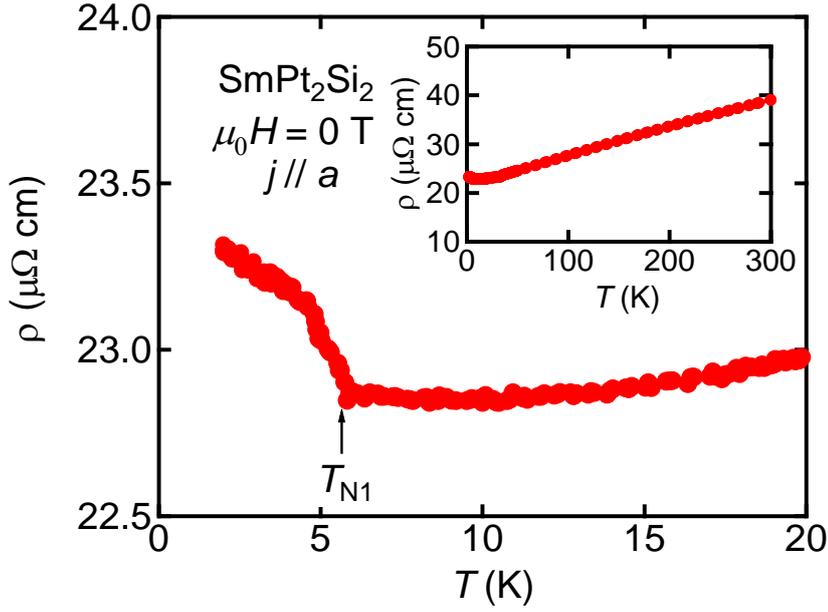

Fig.7 (Color Online) Temperature dependence of resistivity $\rho$ for the current $j // a$.

The pronounced Curie-type $\chi$ observed below 3 K indicates that the phase I is a "partially disordered phase", which includes large amount of paramagnetic (or almost paramagnetic) Sm ions. A Curie-Weiss fitting below 3 K provides $C$= 0.057 emu K/mol and $\theta_p$=-1.34 K, suggesting that 70% of Sm ions are in a paramagnetic state. The realization of such inhomogeneous (probably, spatially modulated) magnetic structures is attributable to geometric frustration in magnetic interactions among Sm magnetic moments. [28,29] Actually, in UPd$_2$Si$_2$, which has the same magnetic-ion configuration in the crystal structure as in SmPt$_2$Si$_2$, the existence of multiple magnetic phases with modulated magnetic structures and similar Curie-type $\chi$ in an ordered phase have been observed;[30] the overall behavior in the *H*-vs-*T* phase diagram can be accounted for on the basis of the axial-next-nearest-neighbor Ising (ANNNI) model taking into account the inherent frustrated magnetic interactions.

In the phase I, the disordered Sm ions can form a Kondo sublattice through the hybridization with conduction electrons, resulting in a quasiparticle mass enhancement. This may account for the largely enhanced $\gamma$ value observed in the phase I. Such partial Kondo lattice formation has been studied theoretically[31] and experimentally in CePdAl,[32,33] and UNi$_4$B[34,35]. Advantages in SmPt$_2$Si$_2$ is that Sm magnetic moments have simple Ising anisotropy and the phase I seems to continue down to $T$=0. Therefore, SmPt$_2$Si$_2$ should provide an appropriate system in which one can study the ground state properties of a "partial Kondo sublattice" immersed in a magnetically ordered phase. With this aim, microscopic measurements in the phase I is in progress.

**Acknowledgment**

We appreciate Ryousuke Shiina for valuable discussions and Wataru Fujita for his technical support in the single-crystal X-ray measurements. This work was supported by a Grant-in-Aid for Scientific Research on Innovative Areas "Heavy Electrons" (No. 20102007) from MEXT and by Grants-in-Aid for Scientific Research C (No. 23540421) and for Young Scientists B (No. 21740261) from JSPS, Japan.